\documentclass[rnote]{aa}
\usepackage{graphicx}
\usepackage[varg]{txfonts}
\usepackage{url}
\usepackage{color}

\begin{document} 

\title{Bernoulli equation and the nonexistence of maximal jets}
\author{Andrzej A. Zdziarski}

\institute{Centrum Astronomiczne im.\ M. Kopernika, Bartycka 18, 00-716 Warszawa, Poland\label{inst1}}

\date{Received 14 December 2015 / Accepted 28 December 2015}

\abstract{
We discuss the idea of maximal jets introduced by Falcke \& Biermann in 1995. According to it, the maximum possible jet power in its internal energy equals the kinetic power in its rest mass. We show this result is incorrect because of an unfortunate algebraic mistake. }

\maketitle

\section{Introduction}
\label{intro}

Relativistic jets are common in AGNs. They accrete black-hole and neutron-star binaries, as well as $\gamma$-ray bursts, and occasionally appear in tidal disruption events. Calculating their properties is obviously of major importance. 

We consider here a result of \citet{fb95}, hereafter FB95. It claims that there is an upper limit to a component of the jet power in its internal energy. This is an influential, highly cited paper. Here, we correct one of its results.

\section{Maximal jets}
\label{maximal}

We follow here the notation of FB95. The jet power, $Q_{\rm jet}$, and its mass flow rate, $\dot M_{\rm jet}$, can be connected to the mass flow rate through the accretion disc, $\dot M_{\rm disc}$, thereby defining two dimensionless constants as in equation (1) of FB95:
\begin{equation}
q_{\rm j}={Q_{\rm jet}\over \dot M_{\rm disc}c^2},\quad q_{\rm m}={\dot M_{\rm jet}\over \dot M_{\rm disc}}.
\label{constants}
\end{equation}
The Bernoulli equation is obtained by dividing the equations of energy and mass conservation. For a conservative jet with the mass dominated by protons, it is given by equation (27) of FB95, which can be expressed as
\begin{equation}
{\gamma_{\rm j}\omega\over n_{\rm tot} m_{\rm p}c^2}= {Q_{\rm jet}\over \dot M_{\rm jet}c^2}={q_{\rm j}\over q_{\rm m}},
\label{b_eq}
\end{equation}
where $\omega$ and $n_{\rm tot}$ are the density of the enthalpy (including rest mass) and of protons, respectively, and $\gamma_{\rm j}$ is the jet bulk Lorentz factor. The enthalpy density for a polytropic gas with the adiabatic index $\Gamma$ is (as in equation 21 of FB95),
\begin{equation}
\omega=m_{\rm p}n_{\rm tot}c^2+\Gamma P_{\rm jet}/(\Gamma-1),
\label{omega}
\end{equation}
where $P_{\rm jet}$ is the jet pressure. Equation (\ref{b_eq}) implies that the maximum possible bulk Lorentz factor, corresponding to the complete conversion of the internal energy into the bulk motion, is $\gamma_{\rm j,max}=q_{\rm j}/ q_{\rm m}$. The sound speed in the jet frame, $\beta_{\rm s}=(\partial P_{\rm jet}/\partial \epsilon)^{1/2}$, where $\epsilon$ is the internal energy density including the rest mass, can be written using the first law of thermodynamics (see, e.g., \citet{konigl80},
\begin{equation}
\beta_{\rm s}^2={\Gamma P_{\rm jet}\over \omega}<\Gamma-1,
\label{sound}
\end{equation}
which is equivalent to equation (23) of FB95, and where the upper limit to the sound speed corresponds to the gas being extremely relativistic, $\omega\gg m_{\rm p}n_{\rm tot}c^2$. Equation (\ref{omega}) can be rewritten as
\begin{equation}
{\omega\over n_{\rm tot}}=m_{\rm p}c^2+{\omega\over n_{\rm tot}}{\beta_{\rm s}^2\over \Gamma-1},
\label{omega2}
\end{equation}
which has the solution of 
\begin{equation}
{\omega\over n_{\rm tot}} ={m_{\rm p}c^2 \over 1-\beta_{\rm s}^2/ (\Gamma-1)},
\label{omega3}
\end{equation}
or equivalently using equation (\ref{b_eq}),
\begin{equation}
\gamma_{\rm j} q_{\rm m}\left[1 \over 1-\beta_{\rm s}^2/ (\Gamma-1)\right]=q_{\rm j}. 
\label{omega4}
\end{equation}
This equation yields the obvious result that the jet gas can be highly relativistic, $\beta_{\rm s}^2\rightarrow \Gamma-1$, only for $q_{\rm m}/q_{\rm j}\ll 1/\gamma_{\rm j}$, i.e., for jets with little relative matter content. On the other hand, $\beta_{\rm s}=0$ corresponds to the jet with the maximum possible bulk Lorentz factor, equal to $\gamma_{\rm j,max}$, after a conversion of all of the internal energy into bulk motion. We note that equation (\ref{omega3}) is identical to equation (A8) of \citet{fukue04} except for the $g_{00}$ metric factor in the latter.

On the other hand, the version of our equation (\ref{omega4}) in FB95, their equation (28), 
\begin{equation}
\gamma_{\rm j} q_{\rm m}\left(1 +{\beta_{\rm s}^2\over \Gamma-1}\right)=q_{\rm j},
\label{omega5}
\end{equation}
is incorrect because of an unfortunate error in the algebraic transformation. A similar (incorrect) form of the Bernoulli equation is also given in equation (1) of \citet*{fmb93}. It appears that the origin of this error is the fact that equation (\ref{omega5}) is the non-relativistic limit of equation (\ref{omega4}), and this would be true if the nonrelativistic form of the sound speed, $\beta_{\rm s}^2=\Gamma P_{\rm jet}/(m_{\rm p}n_{\rm tot}c^2)$, is applied.

As noted by FB95, the maximum value (corresponding to the maximum of $\beta_{\rm s}^2$) of the term in parentheses in equation (\ref{omega5}) is 2. Given that, equation (\ref{b_eq}) implies that the maximum possible total enthalpy density is $\omega=2 n_{\rm tot}m_{\rm p}c^2$; i.e., the maximum internal enthalpy density equals $n_{\rm tot}m_{\rm p}c^2$. Since the jet power equals the enthalpy flux, this implies in turn that the maximum possible component of the jet power in its internal enthalpy equals the jet kinetic power (i.e., in its rest mass). FB95 called this case the ``maximal jet''. However, since equation (\ref{omega5}) is incorrect, this theoretical concept is also incorrect. The correct form of equation (\ref{omega5}) is given by equation (\ref{omega4}), in which the term in brackets can be arbitrarily large, implying no limit on the internal power, hence no maximal jets. We also note that this maximum was supposed to correspond to $\beta_{\rm s}^2=\Gamma-1$, i.e., an extremely relativistic gas, which is not the case for $\omega=2 n_{\rm tot}m_{\rm p}c^2$. This incorrect concept has been adopted in a large number of papers (e.g., \citealt*{falcke96,mff01,markoff03,mn04,mnw05,maitra09,plotkin15}). In particular, the formula (\ref{omega5}) for the maximal jet is one of the underlying assumptions of the complex jet model of \citet{mnw05}, which is widely used to fit broad-band spectra of black-hole binaries and Sgr A*.

FB95 also considered equipartition between magnetic energy density, $u_B$, and the particle internal energy density. If the latter equals the rest mass energy density, we have the magnetization parameter of $\sigma\sim 1/2$, where
\begin{equation}
\sigma\equiv {u_B+p_B\over w},
\label{sigma}
\end{equation}
and $p_B$ is the magnetic pressure. We note that the magnetic enthalpy, $u_B+p_B$, should be included in the Bernoulli equation. This expresses how a jet can be accelerated at the expense of both the magnetic energy and the internal particle energy. Approximately, maximal jets with equipartition correspond to $\sigma\sim 1/2$. This is similar to $\sigma\simeq 1$, up to which the magnetic-to-kinetic energy conversion via differential collimation of poloidal magnetic surfaces is efficient \citep*{tmn09,lyubarsky10,komissarov11}. However, jets can have $\sigma\gg 1$ initially, while blazar jets beyond radio cores have the opening angles $\Theta_{\rm j}\sim (0.1$--$0.2)/\gamma_{\rm j}$ \citep{jorstad05,pushkarev09,clausen13}. Since $\sigma\sim (\Theta_{\rm j}\gamma_{\rm j})^2$ \citep{tmn09,komissarov11}. Some processes should be able to decrease $\sigma$ to values $\ll 1$. 

\section{Conclusion}

We have clarified the issue of the existence of maximal jets, after correcting an algebraic error of FB95. We showed that the maximal internal enthalpy density in jets is not limited by the rest-mass energy density, and correspondingly, the internal jet power is not limited by the (rest-mass) kinetic jet power.

\begin{acknowledgements}
We thank Arieh K{\"o}nigl for valuable comments and for checking the results of this work and Heino Falcke for valuable discussions. This research has been supported in part by the Polish NCN grants 2012/04/M/ST9/00780 and 2013/10/M/ST9/00729.
\end{acknowledgements}

\bibliographystyle{aa}

\end{document}